\documentclass[a4paper, 10pt]{article}
\usepackage[T1]{fontenc}
\usepackage{authblk}
\usepackage[english]{babel}
\usepackage{multicol}
\usepackage[font={footnotesize}]{caption}
\usepackage[margin=.5in]{geometry}
\usepackage{indentfirst}
\usepackage{graphicx}
\usepackage{microtype}
\usepackage{amsthm}
\usepackage{mathtools}
\usepackage{amssymb}
\usepackage{adjustbox}
\usepackage{blindtext}
\usepackage{hyperref}
\usepackage{csquotes}
\usepackage{fancyhdr}
\usepackage{subcaption}
\usepackage{float}
\usepackage{caption}
\usepackage[version=4]{mhchem}
\usepackage[nottoc,notlot,notlof]{tocbibind}
\usepackage{booktabs}
\usepackage{multirow}
\usepackage{lastpage}
\usepackage{times}
\usepackage{setspace}
\usepackage{dblfloatfix}
\usepackage{xcolor}

\title{\bfseries \Large Formation of Artificial Neural Assemblies \\by Biologically Plausible Inhibition Mechanisms\vspace{.5cm}}
\date{}
\author[1]{Lucas Hoff}
\author[1]{Gustavo Soroka}
\author[1]{Mateus Guimarães}
\author[3,4]{Aline Villavicencio}
\author[1,2]{Marco Idiart}

\affil[1]{\small Neuroscience Graduate Program, Federal University of Rio Grande do Sul, Brazil}
\affil[2]{Physics Department, Federal University of Rio Grande do Sul, Brazil}
\affil[3]{Department of Computer Science, University of Exeter, UK}
\affil[4]{Department of Computer Science, University of Sheffield, UK}
\affil[ ]{\texttt {lucas.hoff@ufrgs.br}}
\affil[ ]{\texttt {\{gustavosoroka, mateus.guimaraes048, marco.idiart\}@gmail.com}}
\affil[ ]{\texttt {a.villavicencio@exeter.ac.uk}}

\begin{document}

\maketitle    

\begin{abstract}
As proposed by Hebb’s theory, neural assemblies are groups of excitatory neurons that fire synchronously and exhibit high synaptic density, representing external stimuli and supporting cognitive functions such as language and decision-making. Recently, a model called Assembly Calculus (AC) was proposed, enabling the formation of artificial neural assemblies through the $k$-winners-take-all selection process and Hebbian learning. Although the model is capable of forming assemblies according to Hebb's theory, the adopted selection process does not incorporate essential aspects of biological neural computation, as neural activity, which is often governed by statistical distributions consistent with power-law scaling. Given this limitation, the present work aimed to bring the model's dynamics closer to that observed in real cortical networks. To achieve this, a new selection mechanism inspired by the dynamics of gamma oscillation cycles, called E\%-winners-take-all, was implemented, combined with an inhibition process based on the ratio between excitatory and inhibitory neurons observed in various regions of the cerebral cortex. The results obtained from our model (called E\%-WTA model) were compared with those of the original model, and the analyses demonstrated that the introduced modifications allowed the network's own dynamics to determine the size of the formed assemblies. Furthermore, the recovery rate of these groups, through the evocation of the stimuli that generated them, became superior to that obtained in the original model.
\newline

\end{abstract}

\vspace{.5cm}
\begin{multicols}{2}
    
\section*{Introduction}

According to Hebb, neural assemblies are groups of excitatory neurons exhibiting synchronized firing and recurrent connections among themselves. These groups would be responsible for representing information and generating cognitive functions such as attention, language, and decision-making \cite{hebb_1949}. In his view, neural assemblies would emerge through the continuous and synchronized activation of excitatory neurons, leading to the strengthening of their synapses. Expanding on Hebb’s idea, Moshe Abeles proposed that information propagation in the cerebral cortex would only be possible through the sequential activation of neural groups exhibiting synchronous activity \cite{Abeles_2011}, thus providing further support for the notion that neural assemblies are the functional unit of the nervous system. Experimental evidence also began to support Hebb’s hypothesis. For example, groups of neurons in the visual cortex of mice were observed to be activated both by sensory stimuli \cite{Carrillo-Reid_ensemble2015-gj} and spontaneously \cite{Miller_2014-rs}. Moreover, neural assemblies were artificially generated, and the optogenetic activation of these groups was able to induce behaviors \cite{Carrillo-Reid_impri2016-mc, Carrillo-Reid_control2019-nf}. In the computational theoretical field, models of the hippocampus and cortex proposed by Marr \cite{Marr1970-gx, Marr1971-aq} explained phenomena such as pattern completion through recurrent connections \cite{Yuste_2024}, while the Hopfield model \cite{Hopfield1982-ma} demonstrated that networks with strongly coupled neurons could in principle generate stable firing states, acting as dynamical attractors, which correspond to neural assemblies. Inspired by the hippocampal CA3 region, another model estimated that approximately 225 neurons constitute each memory stored in this associative network \cite{De_Almeida2007-nb}. This finding underscores the necessity of neuronal groups for representing information.

Although considerable advances have been made in the field of neuroscience, particularly from an experimental viewpoint \cite{Yuste2015-calcium,Buzsaki_mtds,twophoton_1990-vx}, and despite our comprehensive understanding of neuronal function at the cellular and molecular levels, we still lack a theoretical framework or formal logic capable of consistently associating observed patterns of neural activity with the emergence of cognitive functions \cite{neuronqa}. With this question in mind, a framework called Assembly Calculus (AC) was proposed, using the neural assembly as its basic unit \cite{christos,random,longterm}. As a first approximation, \cite{christos} proposed a  discrete-time model equipped with a set of operations responsible for the formation and maintenance of neural assemblies. It presents a modular structure of brain areas, whose elements form random synaptic connections. The synaptic weights of each connection are updated using a multiplicative form of Hebbian learning (see Methods), which depends on the sequential activation of two synaptically connected neurons. Inhibition is implemented via a $k$-winners-take-all (k-WTA)  mechanism, which selects the $k$ most strongly stimulated neurons in each area. Building on this basic model, a simple parser was developed \cite{daniel}, and its application to the memorization of temporal sequences was explored \cite{Dabagia_2024}. Additional studies also investigated variations of the original model to examine different functional properties \cite{Constantinides_2021, rec}.

Although the proposed model allows the formation of neural assemblies using simple mechanisms, the k-WTA process may not capture fundamental features of neural computation. For example, in cortical networks, neuronal firing is modulated by oscillatory rhythms such as gamma and theta. The number of neurons that fire within each oscillatory window is not fixed, but instead depends on the distribution and strength of inputs received by each neuron \cite{licurgo}.  By enforcing a fixed number of active neurons, the k-WTA selection process may inadvertently activate weakly stimulated neurons while suppressing others that receive synaptic inputs similar to those of the selected ones. This would not be a concern if the probability of the simultaneous activation of $k$ neurons within a region followed a normal distribution. However, increasing evidence suggests that neuronal firing exhibits power-law distributions \cite{Bialek2015, Bialek2025-kn}.

In this paper, we explore an alternative approach to modeling inhibition in neural networks. Specifically, we use the E\%-winners-take-all (E\%-WTA) mechanism \cite{licurgo}, inspired by the dynamics of gamma oscillations. It has been argued that one of the roles of gamma oscillations is to regulate which neurons fire within each cycle \cite{Fries2007-cx}. In this framework, the number of active neurons is no longer fixed; instead, it depends on the neuron that receives the strongest synaptic input during each iteration. To further enhance network stability, we introduce an auxiliary inhibitory mechanism based on the ratio of excitatory to inhibitory neurons. This addition aims to stabilize the formation and retrieval of neural assemblies. Using this model, we investigate several properties of the resulting assemblies, including their formation dynamics, size distribution, synaptic density, retrieval accuracy, and the degree of overlap between multiple assemblies within the same brain area.




\section*{Methods}
\begin{figure}[H]
\centering
\includegraphics[scale = 0.55]{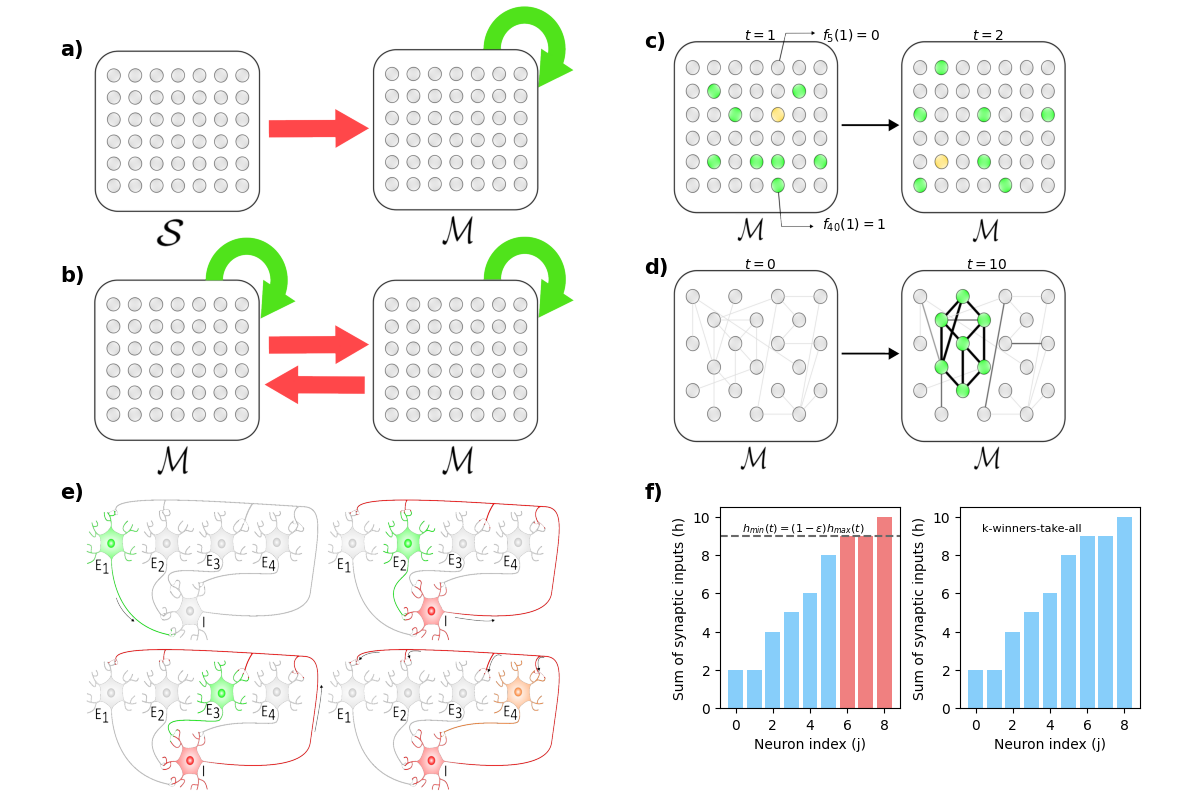}
\caption{Structure and dynamics of the E\%-WTA model. \textbf{a)} Neurons in stimulus areas ($\mathcal{S}$) project only to neurons in memory areas ($\mathcal{M}$) (red arrow), whereas neurons in a memory area can form an internal recurrent network (green arrow), but cannot project back to the stimulus area. \textbf{b)} Neurons in memory areas can project to other memory areas, potentially forming a bidirectional flow of stimuli. \textbf{c)} The firing of a neuron in a memory area $\mathcal{M}$ is represented by the firing state function $(f)$, where $f(t)=1$ for a firing neuron (green), while $f(t)=0$ represents a non-firing neuron (gray). What determines the value of $f(t)$ is the E\%-winners-take-all selection process, which depends on the neuron that received the greatest stimulation (yellow). In each iteration of the model, the neurons that fire can vary, depending on the dynamics of stimuli across the network. \textbf{d)} Synaptic weights are initialized according to Eq. \ref{eq:weights} and adjusted according to Hebb's rule. At the end of the formation process, a neural assembly will be created (green), exhibiting reinforced synapses (dark lines). \textbf{e)} Excitatory neurons are selected to fire during gamma oscillations. The most stimulated neuron fires first ($E_1$), opening a temporal window that allows other excitatory neurons to fire ($E_2$ and $E_3$) until it closes. When the temporal window ends, other excitatory neurons will not fire ($E_4$) because they received little stimulation and are therefore inhibited by inhibitory interneurons ($I$). \textbf{f)} In the original selection process, the number of neurons that fire is fixed, allowing neurons with weak stimulation to fire. In the approach proposed here, the E\%-winners-take-all selection process allows a variable number of neurons to fire (red), but also prevents neurons that received little stimulation from firing (blue). Neurons that fire in each iteration meet the conditions of the E\%-winners-take-all process (dashed line).}
\label{fig:paper_1}
\end{figure}

\noindent
\noindent\textbf{Model architecture.} Following \cite{christos}  we consider a modular network  with $n_A$ modules (artificial brain areas), denoted by  $\mathcal{A}_a$, $a=1,\cdots, n_A$, each one with exactly $n$ neurons, i.e., $|\mathcal{A}_a| = n$.  These areas can be classified as memory areas ($\mathcal{M}$) or as stimulus areas ($\mathcal{S}$)\footnote{Although the original name is ``Sensory Area", we have chosen to use the term ``Stimulus Area" here, since it only serves to generate an external stimulus to form an assembly. Furthermore, the sensory pathways are more complex and are not merely simple systems for transmitting information.} (see Figure \ref{fig:paper_1}a and \ref{fig:paper_1}b)  \cite{lengstein}. Neurons in memory areas form synaptic connections both within their own area and with other memory areas, while neurons in stimulus areas project exclusively to memory areas. Within each memory area, neurons are subject to the same competitive inhibitory process. \newline

\noindent\textbf{Neuronal Dynamic.} The firing of neurons in 
$\mathcal{M}$ occurs in discrete time steps and depends on the total synaptic input they receive at each iteration
\begin{equation}\label{eq:inp}
     h_j(t+1) = \sum\limits_{i=1}^{s} f_i(t) \;\omega_{ij}(t) 
\end{equation}
where  $\omega_{ij}$ is the synaptic 
weight from presynaptic neuron $i$
 to postsynaptic neuron $j$, and $s$ is the number of presynaptic neurons that project to neuron $j$. The function $f_i(t)$ is the firing state \cite{daniel} and indicates if neuron $i$ fired or not at time $t$ (see Figure \ref{fig:paper_1}c). In your model, the firing state is determined by a process called E\%-winners-take-all (E\%-WTA). Matemathically, the firing state function is defined as
\begin{equation}
        f_i(t)=
        \begin{cases}
            1\text{,} &\quad\text{if $i \in F_{t}$ }\\
            0\text{,} &\quad\text{otherwise}
        \end{cases}
        \label{eq2}
    \end{equation}
where $F_{t}$ is the set of all neurons in the area $\mathcal{M}$ that satisfy the E\%-WTA process relationship at iteration $t$. Finally, neurons in the stimulus area represent external stimuli and are therefore set to fire in every iteration. This is implemented by fixing their firing state to 
$f(t)=1$ for all neurons in 
$\mathcal{S}$ that have been selected as part of the stimulus used to form a neural assembly.

Existing connections, in turn, are modified according to a Hebbian-like plasticity rule (see Figure \ref{fig:paper_1}d): 
\begin{equation}\label{eq:heb}
\omega_{ij}(t+1) = \omega_{ij}(t)(1+\beta f_{i}(t)f_{j}(t+1))
\end{equation}
where $\beta$ is the model's learning rate.
\newline

\noindent\textbf{E\%-winners-take-all.} Gamma oscillations consist of a pattern of synchronous neural activity present in the human cortex with a frequency between $30-150$ Hz \cite{buz_2,health}, first observed in the visual cortex of cats \cite{gray} and later in other brain regions \cite{Fries2007-cx}. Their functional role has been associated with various cognitive functions, such as attention, working memory, and episodic memory \cite{Fries2001-rp, Jensen2007-hy, gama_working, Grif}. Although the synchronization of neuronal populations is a fundamental and intrinsic characteristic of gamma oscillations, another property that emerges from the dynamics of the cortical networks that generate them is the ability to select which excitatory neurons should fire in each cycle \cite{Fries2007-cx} (see Figure \ref{fig:paper_1}e). This selection process stems from the interaction between excitatory and inhibitory neurons, where the neuron that receives the greatest stimulation in a given gamma cycle initiates a temporal window during which only a few additional excitatory neurons can fire, before inhibition reaches them. This window is created because the most excited neuron fires first and recruits inhibitory interneurons, which then suppress the activity of the other excitatory neurons in the network. In this way, the temporal window reflects the interval necessary for the inhibition to propagate and reach the other excitatory neurons. One way to describe this selection process associated with inhibitory feedback is through the equation \cite{licurgo}
\begin{equation}
    \frac{E_{max} - E}{E_{max}} \leq \frac{d}{\tau_m} = \epsilon
    \label{emax}
\end{equation}
where $d$ and $\tau_m$ refer, respectively, to the time it takes for the inhibition to reach the excitatory neurons, which begins when the excitatory neuron that received the most stimuli fires, and the membrane time constant.  The equation defines the maximum allowed percentage difference between the total synaptic input (E) of any excitatory neuron and that of the most stimulated neuron ($E_{max}$) for it to fire in the same cycle. If we consider that $E \sim h(t)$, neurons will fire if
\begin{equation}\label{eq:condition}
    \left(1 - \epsilon \right)h_{max}(t) \leq h_{j}(t) \leq h_{max}(t)
\end{equation}
where $h_{max}(t)$ represents the sum of synaptic inputs for the neuron that received the greatest stimulation, and $h_{j}(t)$ represents the sum for a neuron $j$ satisfying the inequality above. Consequently, $F_{t}$\footnote{In the case where $t = 0$, it can be defined that $F_0 = \emptyset$, thereby preventing all neurons in a memory area from being in $F_0$, since $h_t = 0$ satisfies the previously defined equation (Eq. \ref{eq:condition}).} in Eq. \ref{eq2} is defined as 
\begin{equation} 
F_{t} = \left\{ j \in \mathcal{M}\; |\; h_{j}(t) \in \; \left[ \left(1 - \epsilon \right)h_{max}(t) , h_{max}(t) \right] \right\}
\end{equation}

The condition presented here is a biologically plausible alternative to the $k$-WTA process (see Figure \ref{fig:paper_1}f). 
\newline

\end{multicols}

\begin{multicols}{2}

\noindent\textbf{Feedforward inhibition.} The balance between excitatory and inhibitory stimuli is a crucial mechanism for the modulation and formation of neural assemblies \cite{sadra}. Networks that exhibit this control allow external stimuli to be efficiently represented within the network \cite{Zhou2018}, making learning and information storage effective \cite{Sukenik2021}. Structurally, this balance is maintained by an asymmetric distribution of excitatory and inhibitory neurons, a pattern observed across various cortical regions \cite{Alreja2022}.

To incorporate this fundamental aspect into the model architecture, we assume that each existing connection has a probability $p_i$ of being inhibitory. Therefore, if the probability of a synaptic connection between two neurons is $p_s$, in the model the initial synaptic weights are assigned as positive or negative according to the following expression
\begin{equation}\label{eq:weights}
 \omega_{ij} =\begin{cases}
       \;\;\quad1\text{,} &\quad\text{$p_s(1-p_i)$}\\
\;\;\;\;\omega_{inh}\text{,} &\quad\text{$p_sp_i$}\\\;\;\quad0\text{,} & \quad\text{$1-p_s$}
     \end{cases}
\end{equation}
where $\omega_{inh} <0$ and $\omega_{ij} = 0$ indicates that there is no synapse from $i$ to $j$. 
\newline

\noindent\textbf{Conditions for assembly formation.} In the formulation proposed by \cite{christos}, neural assemblies are formed through successive firings in area $\mathcal{M}$ due to  activity in area $\mathcal{S}$ (see Figure \ref{fig:paper_3}a). The condition for the assembly to be considered formed is that there are no new winners in a given iteration. The number of new winners ($N_t$) is defined as the number of neurons that fired for the first time at time $t$. Therefore an assembly is formed if $N_t = 0$. Since only $k$ neurons are allowed to fire in each iteration, the assemblies will have size $k$. However, even if there are no more winners, nothing guarantees that the $k$ neurons that fired in the last iteration are the same as those that fired in the previous iteration (see Figure \ref{fig:paper_3}b). Because of this, in our model we consider that an assembly is a stationary state of the dynamics with extra conditions: 

\begin{enumerate}
    \item[i)] {\bf Stationary pattern}. 
The number of neurons active at time $t$ must be equal to the number active at the next iteration: 
$$|F_t| = |F_{t+1}|$$
$$\sum_i^{|\mathcal{M}| } f_i(t) = \sum_j^{|\mathcal{M}| }f_j(t+1)$$
\begin{equation}\label{eq:stat}
    \sum_i^{|\mathcal{M}| } \Delta f_i = 0
\end{equation}
where $\Delta f_i = f_i(t+1) - f_i(t)$.

\item[ii)]{\bf Synchronization}.
If the set $X_{t}$ represents the neurons that did not fire in iteration $t-1$ but fired in iteration $t$, then to ensure that the neurons firing remain the same, we must have:
\begin{equation}\label{eq:sync}
    X_{t} = \emptyset
\end{equation}
\item[iii)]{\bf Higher synaptic density}. Neural assemblies should have a higher synaptic density relative to the area they are located
\begin{equation}
    D_{A} > D_{\mathcal{M}}
\end{equation}
with $D_{A}$ representing the density of assembly $A$, while $D_{\mathcal{M}}(= p_s)$ is the synaptic density of the area $\mathcal{M}$, considering only the connections between its own neurons. Since we are dealing with random graphs, we can measure the synaptic density as:
\begin{equation}
    D = \frac{|S|}{|N|(|N| - 1)}
\end{equation}
where $|S|$ represents the number of synapses formed between the neurons of a group, while $|N|$ represents the number of neurons in that group.
\end{enumerate}

Although the conditions for the formation of an assembly differ from those of the original model, the dynamics across iterations are similar. Initially, at time $t = 0$, only $k_{s}$ neurons from the sensory area $\mathcal{S}$ fire. At the subsequent time, $t = 1$ the stimulus from $\mathcal{S}$ cause $\left|F_{1}\right|$ neurons in $\mathcal{M}$ to fire. The neurons that  fire are those whose synaptic input satisfies the E\%-WTA rule. Since no neuron in  $\mathcal{M}$ fired during the previous iteration ($|F_{0}| = 0$) then $X_{1} = F_{1}$. Additionally, at $t = 1$, the $k_{s}$ neurons that fired in $\mathcal{S}$ during the previous iteration do not only fire again now but continue to fire throughout the entire process, as an assembly is formed by successive firing. At time $t = 2$ the area $\mathcal{M}$ receive stimuli both from $\mathcal{S}$ and from the neurons that fired in $\mathcal{M}$ during the previous iteration ($F_{1}$). As in the previous iteration, $\left|F_{2}\right|$ neurons fire, and $X_{2} = F_{2} - F_{1}$. This process continues until the first two conditions mentioned are met. If the set of neurons to which the process has converged satisfies the third condition, an assembly is considered to be formed.\newline

\noindent\textbf{Model simulations and parameters.} All simulations involved two artificial brain areas, a stimulus area and a memory area (see Fig \ref{fig:paper_1}a), each containing $10^3$
neurons. Synaptic connections were formed with a probability 
$p_s=0.5$, and with 
$p_i$ = 0.2 chance of being inhibitory \cite{Sahara2012}. In the original model simulations, the typical probability for a synapse to be formed was maintained ($p_s = 0.1$). The stimuli in $\mathcal{S}$  consisted of $k_s = 200$ randomly selected neurons, whereas in the original model, $k_s = 37\text{ and } 200$ neurons were randomly selected. To evaluate the characteristics of the neural assemblies, several simulations were carried out. In each simulation, new memory and stimulus areas were created so that only one neural assembly was stored in the memory area. Thus, in each simulation, the areas were reinitialized, the synaptic connections re-established, and the set of stimuli re-selected, keeping the system always in its initial state at the beginning of each simulation. In the case of studying the overlap of multiple neural assemblies in the same memory area, the system was restarted after the formation of $10$ consecutive neural assemblies. Using this process, we were able to obtain the distributions of several features of the formed assemblies for different synaptic plasticity values. The number of iterations required for the formation of a neural assembly, the percentage of neurons recovered from a formed neural assembly, and the overlap of these groups were compared with the results obtained in the original model. Regarding the size, synaptic density, and formation process in the model, these were evaluated only between different synaptic plasticity values. This is because the size of the neural assemblies is no longer fixed, the basal synaptic density is different (as different connection probabilities were used), and the neural assembly formation process occurred in a distinct manner (see Results). Finally, the parameters $\tau_m$ and $d$ were set to the same values as in \cite{licurgo}, 30 ms and 3 ms, respectively. For a summary of all the parameters used, refer to the Table~\ref{tab:tb_2}. For the characteristics of neural assemblies the Shapiro-Wilk test was applied to check for data normality; non-normally distributed data are presented using medians and non-parametric tests were used to compare distributions between two or more groups. The code for the simulations and plots, as well as the data files, can be found here (\href{https://github.com/Solution-Epsilon/AC_Emax}{Github}) and the simulations and the statistical analyses were conducted using Python.
\end{multicols}

\begin{table}[h]
\begin{adjustbox}{center}
\begin{tabular}{llcc}
\hline
\multicolumn{1}{c}{\multirow{2}{*}{Parameters}} & \multicolumn{1}{c}{\multirow{2}{*}{Description}}          & \multicolumn{2}{c}{Value}      \\ \cline{3-4} 
\multicolumn{1}{c}{}                            & \multicolumn{1}{c}{}                                      & E\%-WTA model & AC model       \\ \hline
$n$                                             & Number of neurons per area                                & $1000$        & $1000$         \\
$n_{A}$                                         & Number of areas                                           & $2$           & $2$            \\
$p_{s}$                                         & Probability of a synapse being formed between two neurons & $0.5$         & $0.1$          \\
$p_{i}$                                         & Probability of a synapse being inhibitory                 & $0.2$         & N/A            \\
$\tau_{m}$                                      & Membrane time constant (ms)                               & $30$          & N/A            \\
$d$                                             & Delay time to reach a group of excitatory neurons (ms)    & $3$           & N/A            \\
$\beta$                                         & Synaptic plasticity of the model                          & $0.001 - 0.1$ & $0.001 - 0.1$  \\
$k_{s}$                                         & Stimulus size used (number of neurons)                    & $200$         & $37$ and $200$ \\
$k$                                             & Number of neurons that can fire in each memory area       & N/A           & $37$           \\
$\omega_{inh}$*                                 & Inhibitory synaptic weight                                & $-0.2$        & N/A            \\ \hline
\end{tabular}
\end{adjustbox}
\caption{Parameters for each of the models.}
\label{tab:tb_2}
\end{table}

\begin{multicols}{2}

\section*{Results}

\begin{figure}[H]
\centering
\includegraphics[width=1\linewidth]{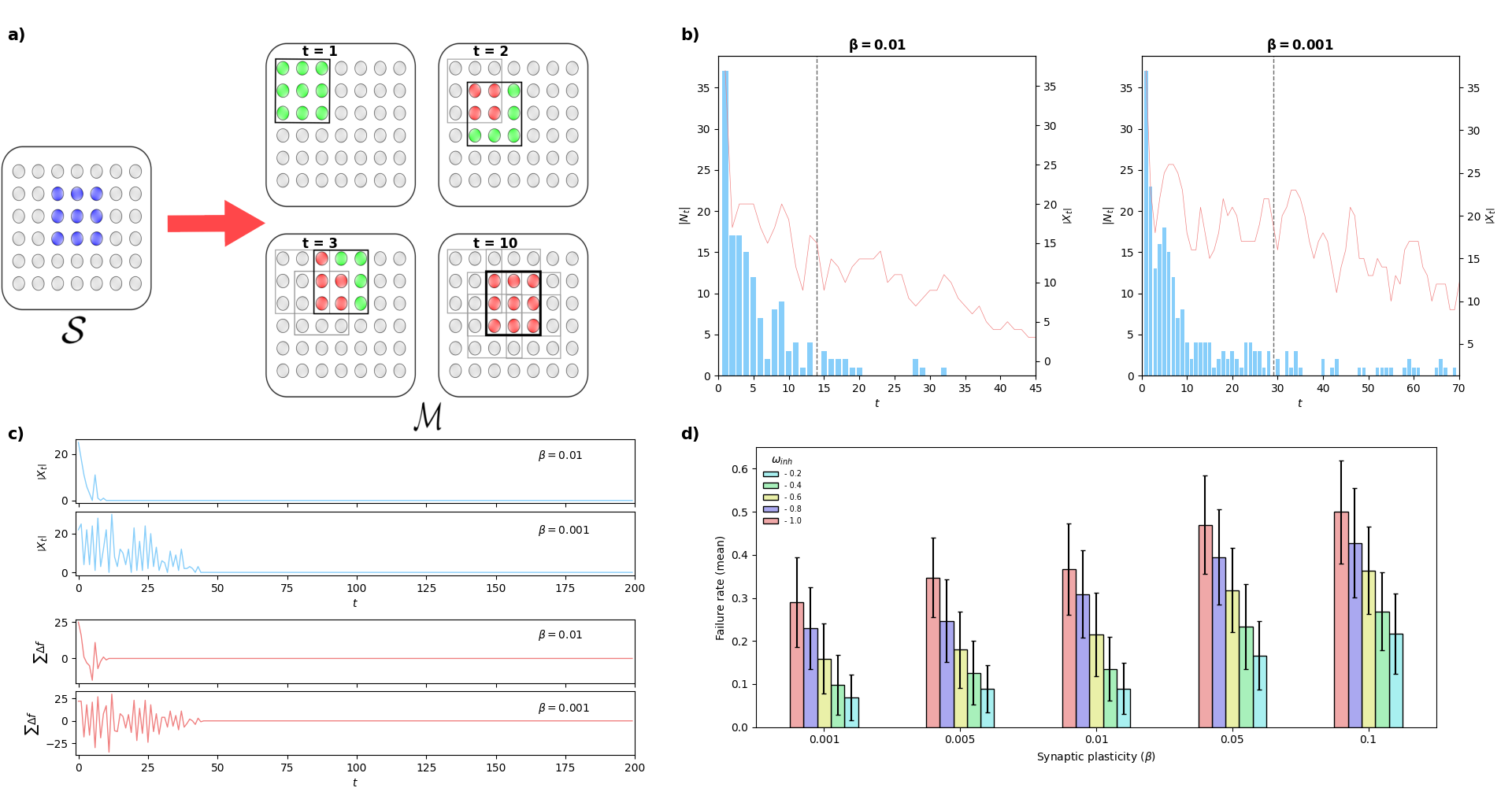}
\caption{Neural assembly formation in AC and E\%-WTA models. \textbf{a)} Neural assemblies in the AC model form through the successive firing of a set of neurons in $\mathcal{S}$ (blue). Over iterations, new neurons fire for the first time (green), while other neurons might fire in multiple iterations (red). The condition for formation is that there are no new neurons ($|N_{10}| = 0$ ). \textbf{b)} AC model behavior during formation ($k_s = 37$). As iterations progress ($t$ - horizontal axis), the number of new neurons decreases ($|N_t|$ - left vertical axis, blue). For a certain value of $t$, $|N_t| = 0$ (vertical dashed line), indicating no more new neurons. When this occurs in the original model, a neural assembly is considered formed. However, if iterations continue, the neurons that fire won't always be those belonging to the formed assembly ($|X_t|$ - right vertical axis, red). Thus, even when the formation condition is satisfied, there is no guarantee that the assembly neurons will continue to fire synchronously as iterations progress. \textbf{c)} Behavior of the first two conditions in the E\%-WTA model. In all simulations performed in this study, the first two conditions converged to zero. \textbf{d)} Failure rate (mean) of neural assembly formation in the E\%-WTA model (normalized by the number of simulations for each parameter tested) as a function of synaptic plasticity ($\beta$) for different feedforward inhibition values ($\omega_{inh}$), considering a baseline size ($|A|_{min}$) of 6 neurons. After 100 simulations for each parameter ($\beta$ and $\omega_{inh}$), the best values for neural assembly formation in your model are: $\omega_{inh} = -0.2$ and $\beta \leq 0.01$. (Parameters: see Table~\ref{tab:tb_2})}
\label{fig:paper_3}
\end{figure}

\noindent\textbf{Formation conditions in the E\%-WTA model.} Initially, the first two formation conditions defined (equations \eqref{eq:stat} and \eqref{eq:sync}) were evaluated. It was observed that, across all simulations performed for distinct values of synaptic plasticity, both conditions converged to zero (see Figure \ref{fig:paper_3}c). This suggests that the initial two conditions can serve as a foundation for the formation of assemblies within the E\%-WTA model. Even though the first two conditions are consistently satisfied for the defined parameters, this fact won't imply the formation of a neural assembly. As previously stipulated, an assembly is formed only when all imposed conditions are fulfilled. However, it was observed that some of the formed groups exhibited synaptic density lower than that established for the memory area ($D < D_{\mathcal{M}}$). Furthermore, groups composed of merely one or two neurons were also noted. Consequently, while the first two conditions may be satisfied, failures associated with synaptic density and the size of the assemblies can emerge. To address these problems, several simulations were conducted for different values of synaptic plasticity ($\beta$) and inhibitory synaptic weight ($\omega_{inh}$). The purpose was to identify the optimal values for the model, thereby making it less susceptible to failures during the formation process. Furthermore, a basal value ($|A|_{min}$) of $6$ neurons was established as the minimum size for a neuronal group to be considered an assembly, thus avoiding groups containing only a single neuron. From these considerations, a reduced percentage of failures was observed with the decrease in both synaptic plasticity and feedforward inhibition values (see Figure \ref{fig:paper_3}d). Based on these findings, it was concluded that the most suitable values for the formation of a neural assembly in the E\%-WTA model are: $\omega_{inh} = - 0.2$ and $\beta \leq 0.01$.\newline
\end{multicols}

\begin{multicols}{2}
\noindent\textbf{Number of iteration, size and synaptic density.} Considering initially the E\%-WTA model, the analysis showed a statistically significant difference in the number of iterations ($T$) for different values of synaptic plasticity ($\beta$), where it was observed that the median [IQR]\footnote{IQR: Interquartile range.} number of iterations for $\beta = 0.001$ ($64 \left[44 - 83\right]$) was higher than the median for the other synaptic plasticity values (see Table \ref{tab:table_2}) (Kruskal-Wallis test with Dunn’s post hoc, $p \leq 0.001$). When the number of iterations was compared between the two models (E\%-WTA and AC), the analysis showed a statistically significant difference, with the most substantial median difference occurring at $\beta = 0.001$, where the E\%-WTA model presented a median [IQR] of iterations ($64\left[44-83\right]$) higher than that of the AC model ($19\left[15-24\right]$). However, for other values of synaptic plasticity, the median number of iterations was higher in the AC model (Mann-Whitney test, $p \leq 0.001$). Therefore, synaptic plasticity influences the number of iterations required for the formation of neural assemblies in the E\%-WTA model, indicating that lower levels of synaptic plasticity may lead to a slower formation process. Additionally, the formation process in the original model was slightly slower than in the E\%-WTA model, suggesting that the approach proposed here may accelerate the formation process. 

\begin{figure}[H]
\centering
\includegraphics[width=1\linewidth]{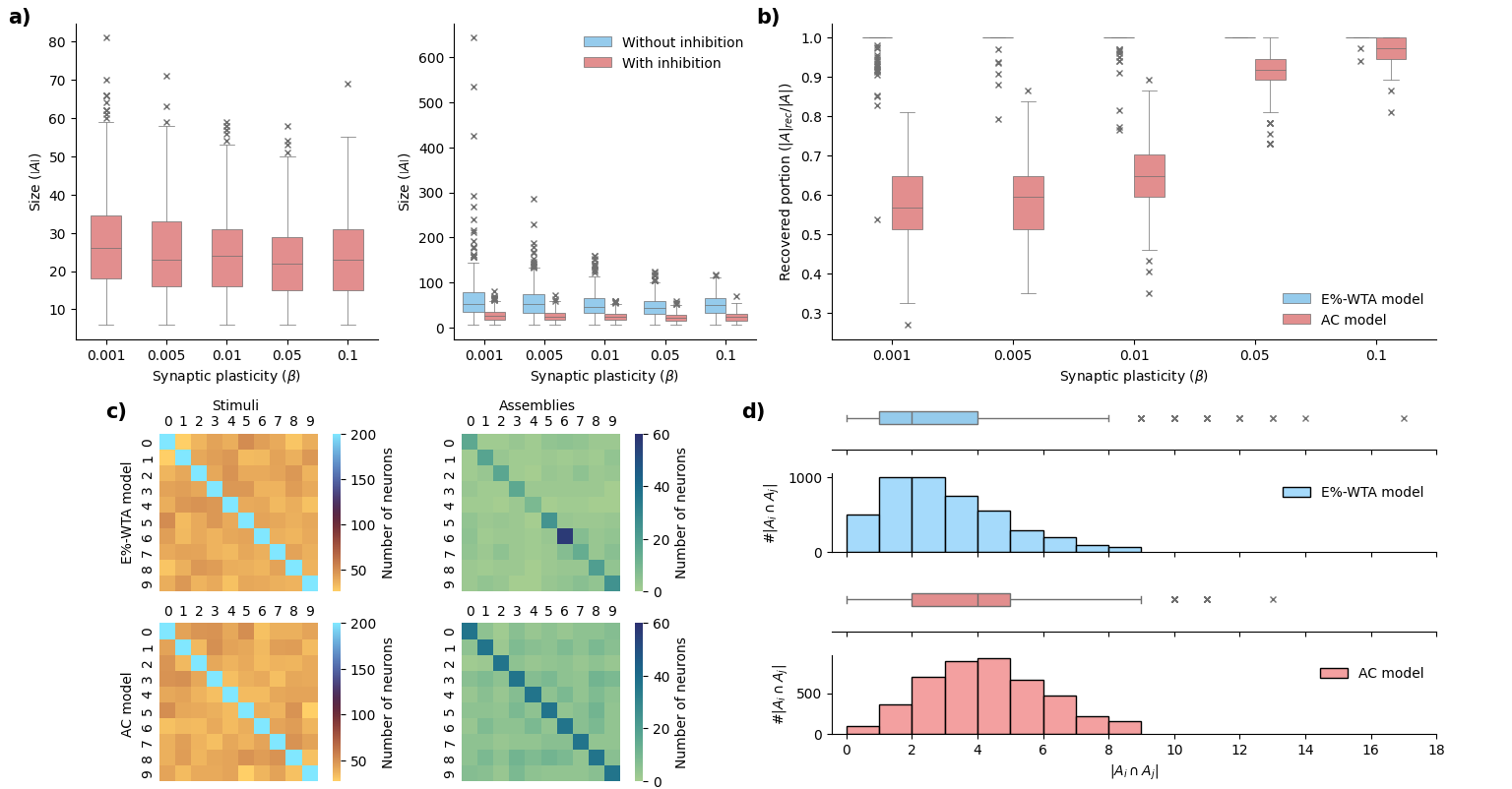}
\caption{Characteristics of the assemblies formed in the AC and E\%-WTA models as a function of synaptic plasticity. \textbf{a)} Size of the neural assemblies in the E\%-WTA model with and without (only positive synaptic weights) feedforward inhibition. \textbf{b)} Retrieval of the formed assemblies, where recovered portion ($|A|_{rec}$) was normalized by the size ($|A|$). \textbf{c)} Overlap matrix of the stimuli (right) and of the assemblies in a memory area (left). E\%-WTA model – Top matrices; AC model – Bottom matrices. The diagonal of the matrices represents the size of the neural assemblies and the size of the stimuli. \textbf{d)} Distribution of the overlap ($\#|A_i\cap A_j|$) between the models ($\beta = 0.01$ and $k_s = 200$ for both models). Crosses ($\times$) represent the outliers of the presented distributions.}
\label{fig:paper_4}
\end{figure}

As previously discussed, the E\%-winners-take-all selection method does not provide control over the number of neurons that will fire in each iteration. As a result, we have no control over the size of the neural assemblies formed, meaning that some assemblies may include a number of neurons comparable to the total number of neurons in the area. As a control measure, an feedforward inhibition method was implemented based on the excitation-inhibition balance present in cortical networks (see Methods). The analysis showed a statistically significant difference in the size of neural assemblies ($|A|$) formed with and without (only positive synaptic weights) feedforward inhibition in the E\%-WTA model (see Figure \ref{fig:paper_4}a). It was observed that the largest medians, for each synaptic plasticity, were obtained in the model without the inhibition method (see Table \ref{tab:table_2}), with the highest median [IQR] recorded for synaptic plasticity $\beta = 0.001$ ($52\left[35 - 79\right]$) (Mann-Whitney test, $p \leq 0.001$). Regarding sizes, when considering only the E\%-WTA model with feedforward inhibition, a statistically significant difference was observed only between synaptic plasticity $\beta = 0.001$, with a median [IQR] of $26\left[18-34\right]$, and $\beta = 0.1$ and $0.05$, with medians of $23\left[15-31\right]$ and $22\left[15-29\right]$, respectively (Kruskal-Wallis test with Dunn’s post hoc, $p \leq 0.001$). Therefore, based on the analysis, we can conclude that the inhibition method applied effectively controlled the size of the neural assemblies formed, indicating that E\%-WTA model enables efficient storage of an external stimulus in a memory area by regulating the size of the neural assemblies. Furthermore, the size of neural assemblies appears to be minimally influenced by changes in synaptic plasticity values.

Another feature analyzed was the synaptic density of the neural assemblies formed, considering only the model with inhibition. As previously defined, a group of neurons is considered a neural assembly if its synaptic density is greater than the synaptic density of the memory area, where $D_{\mathcal{M}} = p_s$. The analysis showed a statistically significant difference between the synaptic densities obtained for $\beta = 0.001, 0.005$, and $0.01$ and those obtained for $\beta = 0.05$ and $0.1$ (see Table \ref{tab:table_2}), with the highest median [IQR] observed for synaptic plasticity $\beta = 0.005$ ($0.555\left[0.540 - 0.570\right]$) (Kruskal-Wallis test with Dunn’s post hoc, $p \leq 0.001$).
Therefore, based on the analysis, we can conclude that the model is capable of generating neural assemblies with synaptic densities higher than that of the memory area. However, synaptic plasticity in the model showed a slight influence on this value, indicating a significant increase in density only for lower synaptic plasticity values. \newline

\noindent\textbf{Memory retrieval and overlap of multiples assemblies.} Another characteristic evaluated between the E\%-WTA and AC models was the ability to recover a neural assembly after its formation, in the presence of the same stimulus that originally generated it (see Figure \ref{fig:paper_4}b). Based on the obtained data, there was a statistically significant difference between the processes (see Table \ref{tab:table_2}), where for all tested synaptic plasticity levels, the medians for the E\%-WTA model were higher than those of the AC model in terms of recovery of these neural groups (Mann-Whitney test, $p \leq 0.001$). Moreover, when comparing different synaptic plasticity levels within the E\%-WTA model, the tests indicated a significant statistical difference among the data; however, the medians remained the same (Kruskal-Wallis test with Dunn’s post hoc, $p \leq 0.001$). Therefore, we can conclude that the adaptations made to the original model enable superior recovery of the neural assemblies formed in a memory area through the reactivation of the same stimulus that generated them. Furthermore, although the tests revealed statistical differences in the data for each synaptic plasticity level, this did not prevent the recovery of a substantial portion of the formed neural assemblies. 

Finally, the overlap between neural assemblies was analyzed by considering the formation of multiple assemblies within the same memory area, using stimuli of the same size ($k_s = 200$ neurons) (see Figure \ref{fig:paper_4}c and \ref{fig:paper_4}d). Initially, an overlap matrix was defined. The overlap matrix is a symmetric matrix in which the intersection of a row and a column indicates the intersection between two neural assemblies within the same area, that is, $\alpha_{ij} = |A_{i} \cap A_{j}|$, whereas its diagonal represents the sizes of the assemblies ($\alpha_{ij} = |A_{i}|$ for $i = j$). From the analysis between the two models, a statistically significant difference was observed, where the AC model showed a higher median [IQR] overlap ($4\left[3-6\right]$) compared to the E\%-WTA model ($2\left[1-4\right]$) (Mann-Whitney test, $p \leq 0.001$). Therefore, we see that the E\%-WTA model better orthogonalizes the formed neural assemblies, even in the presence of considerable overlap between the stimuli used.
\end{multicols}
\begin{table}[h]
\begin{adjustbox}{width=\columnwidth,center}
\begin{tabular}{llcccccl}
\hline
\multicolumn{2}{l}{\multirow{2}{*}{}}                       & \multicolumn{5}{c}{\textbf{Synaptic plasticity ($\beta$)}}                                                                                                                                                                                                               & \multirow{2}{*}{\textbf{*p-value}} \\ \cline{3-7}
\multicolumn{2}{l}{}                                        & 0.1                                                 & 0.05                                                & 0.01                                                & 0.005                                               & 0.001                                               &                                    \\ \hline
\textbf{Number of iterations ($T$)} &                     &                           &                           &                        &                        &                        &                                    \\
                                   & E\%-WTA model & $4\left[3-5\right]^{a} (81.6\%)$    & $4\left[4-5\right]^{b}(79.2\%)$    & $10\left[8-11\right]^{c}(89.8\%)$ & $16\left[12-20\right]^{d}(92.4\%)$ & $\boldsymbol{64\left[44-83\right]}^{e}(92.6\%)$ & \multicolumn{1}{c}{$\leq 0.001$}             \\
                                   & AC model   & $\boldsymbol{6 \left[6 - 7\right]}$ & $\boldsymbol{8 \left[7 - 9\right]}$ & $\boldsymbol{17\left[14-20\right]}$ & $\boldsymbol{20\left[15-24\right]}$ & $19\left[15-24\right]$ & \multicolumn{1}{c}{N/A}            \\
                                   & \textbf{**p-value}   & $\leq 0.001$                        & $\leq 0.001$                         & $\leq 0.001$                     & $\leq 0.001$                      & $\leq 0.001$                    & \multicolumn{1}{c}{}               \\
\textbf{Size ($|A|$)}               &                    &                                                     &                                                     &                                                     &                                                     &                                                     &                                    \\
                                       & With feedforward inhibition     & $23\left[15-31\right]^{a}(81.6\%)$                              & $22\left[15-29\right]^{a}(79.2\%)$                              & $24\left[16-31\right]^{ab}(89.8\%)$                              & $23\left[16-33\right]^{ab}(92.4\%)$                              & $26\left[18-34\right]^{b}(92.6\%)$                              & \multicolumn{1}{c}{$\leq 0.001$}             \\
                                       & Without feedforward inhibition  & $\boldsymbol{49 \left[33 - 66\right]}(92.2\%)$                           & $\boldsymbol{44\left[29 - 59\right]}(96.6\%)$                           & $\boldsymbol{46\left[32-65\right]}(99.2\%)$                              & $\boldsymbol{51\left[33-73\right]}(97.8\%)$                              & $\boldsymbol{52\left[35-79\right]}(99.4\%)$                              & \multicolumn{1}{c}{N/A}             \\
                                       & \textbf{**p-value}  & $\leq 0.001$                                                  & $\leq 0.001$                                                   & $\leq 0.001$                                                  & $\leq 0.001$                                                   & $\leq 0.001$                                                 & \multicolumn{1}{c}{}               \\
\textbf{Synaptic density ($D_{A}$)}   &                    & \multicolumn{1}{l}{}                                & \multicolumn{1}{l}{}                                & \multicolumn{1}{l}{}                                & \multicolumn{1}{l}{}                                & \multicolumn{1}{l}{}                                &                                    \\
                                       &                    & \multicolumn{1}{l}{$0.534\left[0.518-0.554\right]^{a}(81.6\%)$} & \multicolumn{1}{l}{$0.541\left[0.525-0.561\right]^{b}(79.2\%)$} & \multicolumn{1}{l}{$0.550\left[0.536-0.567\right]^{c}(89.8\%)$} & \multicolumn{1}{l}{$0.555\left[0.540-0.570\right]^{c}(92.4\%)$} & \multicolumn{1}{l}{$0.552\left[0.539-0.567\right]^{c}(92.6\%)$} & \multicolumn{1}{c}{$\leq 0.001$}             \\
                                       & \textbf{}          & \multicolumn{1}{l}{}                                & \multicolumn{1}{l}{}                                & \multicolumn{1}{l}{}                                & \multicolumn{1}{l}{}                                & \multicolumn{1}{l}{}                                &                                    \\
\textbf{Recovered portion ($|A|_{rec}/|A|$)} &                               &                           &                           &                        &                        &                                            &                                    \\
                                & E\%-WTA model & $\boldsymbol{1.00\left[1.00-1.00\right]}^{abd}$    & $\boldsymbol{1.00\left[1.00-1.00\right]}^{d}$    & $\boldsymbol{1.00\left[1.00-1.00\right]}^{abe}$ & $\boldsymbol{1.00\left[1.00-1.00\right]}^{ad}$ & \multicolumn{1}{c}{$\boldsymbol{1.00\left[1.00-1.00\right]}^{e}$} & \multicolumn{1}{c}{$\leq 0.001$}             \\
                                & AC model    & $0.97 \left[0.94 - 1.00\right]$ & $0.91 \left[0.89 - 0.94\right]$ & $0.64\left[0.59-0.70\right]$ & $0.59\left[0.51-0.64\right]$ & \multicolumn{1}{c}{$0.56\left[0.51-0.64\right]$} & \multicolumn{1}{c}{N/A}             \\
                                & \textbf{**p-value}            & $\leq 0.001$                        & $\leq 0.001$                         & $\leq 0.001$                     & $\leq 0.001$                      & $\leq 0.001$                                        & \multicolumn{1}{c}{}               \\ \hline
\end{tabular}
\end{adjustbox}
\caption{Several characteristics of the assemblies as a function of synaptic plasticity ($\beta$) across 500 simulations. The size ($|A|$) was evaluated with and without (only positive synaptic weights) feedforward inhibition. For the retrieval of an assembly ($|A|_{rec}/|A|$) 200 simulations were performed. Data for each distribution are presented as median and interquartile range (25th and 75th percentiles). The percentage next to the medians indicates the proportion of the 500 simulations that formed a neural assembly and were therefore considered in the analysis and statistical tests. For the AC model, this is not taken into account because the original conditions guarantee the formation of the neural assembly. In the retrieval of a neural assembly, the percentage is not considered because the analysis is performed after the formation. Bolded median indicates the highest median value between two models or between E\%-WTA model with or without feedforward inhibition. Different superscript letters indicate statistically significant differences between data for each synaptic plasticity level. *Kruskal-Wallis test with Dunn's post hoc. **Mann-Whitney test. N/A --- Not applicable. Significance level was set at 5\% for all analyses.
\label{tab:table_2}}
\end{table}
\begin{multicols}{2}
    
\begin{figure}[H]
\centering
\includegraphics[width=1\linewidth]{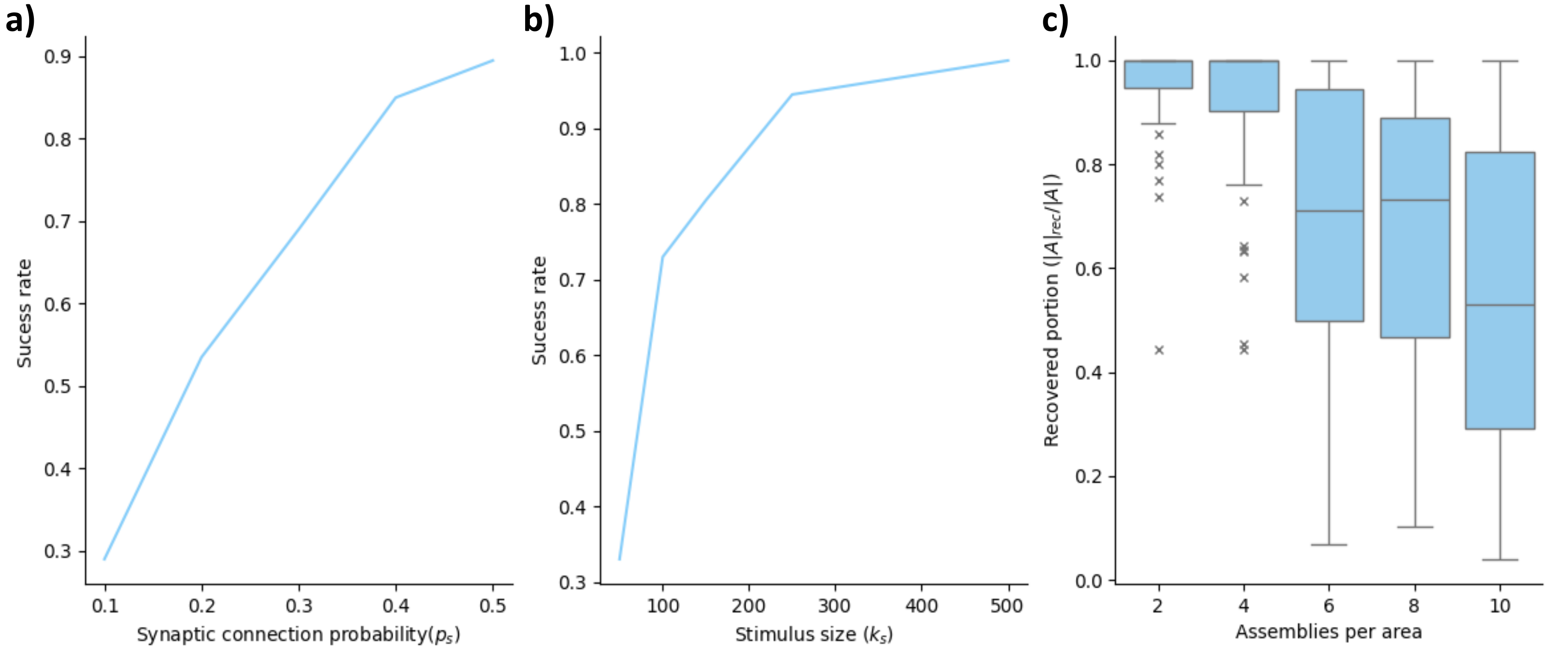}
\caption{ E\%-WTA model behavior as a function of \textbf{a)} the probability of synaptic connection ($p_s$) and \textbf{b)} the stimulus size ($k_s$) (200 simulations for each value of $p_s$ and $k_s$ with $\beta = 0.01$). Lower values lead to lower values of success rate, thereby limiting the applicability of operations defined in the original model. The success rate (normalized by the number of simulations) is defined here as the ratio between the number of simulations that formed a neural assembly and the total number of simulations. \textbf{c)} The formation of more than one neural assembly in a memory area allows only partial retrieval of the first neural assembly formed (50 simulations performed in each case with $\beta = 0.01$).}
\label{fig:paper_5}
\end{figure}

\noindent\textbf{Limitations.} Although the results presented in this study show that the E\%-WTA model outperforms the original model in several aspects, it exhibits limitations when changes are made to the values of the synaptic connection probability ($p_s$) and the stimulus size ($k_s$) (see Figure \ref{fig:paper_5}a and \ref{fig:paper_5}b). For values lower than those used, the E\%-WTA model tends to fail in forming neural assemblies, either due to synaptic densities lower than that of the memory area or by forming groups with too few neurons that no satisfy the third condition. As a consequence, because smaller stimuli make the model more prone to failure, it becomes unfeasible to apply the pre-defined operations from the original model \cite{christos}. Another limitation of the E\%-WTA model is that neural assemblies that are formed may end up being destroyed by the formation of others in the same memory area, allowing only partial retrieval (see Figure \ref{fig:paper_5}c).

\section*{Discussion}

Originally, the Assembly Calculus model consists of an artificial neural network with randomly assigned synapses of positive synaptic weight, capable of forming neural assemblies through synaptic learning as proposed by Donald O. Hebb \cite{hebb_1949} and the $k$-winners-take-all selection process \cite{christos}. Considering that the AC model does not fully capture certain biological aspects of neural computation, the main objective of this work was to replace the original selection process with two biologically plausible mechanisms: the first based on the dynamics of gamma oscillations, called E\%-winners-take-all, and the second based on the ratio between excitatory and inhibitory neurons present in the cerebral cortex, allowing for the existence of synapses with negative weights. Although in the original model and in Hebb’s proposal neural assemblies are formed considering only excitatory synaptic interactions \cite{Traub2020-do}, inhibition plays a fundamental role in the formation of these groups of neurons \cite{sadra}. Thus, the addition of negative synapses enabled inhibitory effects in the summation of synaptic signals received by an artificial neuron, modulating the formation of neural assemblies and, consequently, the size of these groups.

With the implemented adaptations, the formation of neural assemblies was evaluated using new formation criteria based on typical characteristics of these groups: size, synchrony, and synaptic density. The results indicated that, in all simulations, the conditions related to size and synchrony were consistently met, regardless of synaptic plasticity. However, the condition related to synaptic density was not always fulfilled, suggesting potential limitations of the model in forming neural assemblies. Additionally, it was necessary to establish an arbitrary minimum threshold for a group of neurons to be considered a neural assembly, thereby avoiding the inclusion of groups consisting of only one or two neurons. Following this, the formation time, size, synaptic density, and recovery capacity of the neural assemblies were analyzed. It was observed that the E\%-WTA model exhibited shorter formation times compared to the original model for most tested levels of synaptic plasticity. The inhibitory mechanism allowed for greater control over assembly size, preventing representations of external stimuli from occupying large portions of a memory area. This result is consistent with theoretical findings, which indicate that inhibition assists in the maintenance of memory storage \cite{Mongillo2018-ux}. Both size and synaptic density showed low sensitivity to variations in synaptic plasticity, indicating greater stability of the E\%-WTA model across different learning rates. Regarding recovery, the E\%-WTA model was able to recall nearly all previously formed assemblies, whereas the original model performed less effectively, with strong dependence on synaptic plasticity levels. Finally, in the evaluation of multiple assembly overlaps within the same memory area, the E\%-WTA model demonstrated a greater capacity for orthogonalization, even in the presence of significant overlap between distinct stimuli used in the formation of these groups.

Based on the obtained results, it is concluded that the E\%-WTA model was capable of forming neural assemblies of different sizes and fully recovering them across a wide range of synaptic plasticity values. Despite the discussed limitations, the inclusion of the two proposed mechanisms enabled the model to modulate the internal representations of stimuli generated in a stimulus area. These findings reinforce the hypothesis that a biologically plausible process for selecting neurons to fire, combined with the balance between excitation and inhibition, is essential for the efficient representation of information in cortical networks.

Other characteristics not addressed in this work may be explored in future studies. One example is the phenomenon of pattern completion, which refers to the ability to retrieve a neural assembly from the activation of only a fraction of its neurons \cite{Yuste_2024}. Furthermore, it would be of interest to investigate the possibility of implementing the operations defined in the original model \cite{christos}, aiming to make the model more dynamic and applicable across different computational contexts. Another fundamental aspect that deserves attention is the model’s ability to retrieve assemblies after the formation of multiple others. It is well known that artificial neural networks tend to suffer from catastrophic forgetting \cite{Kirkpatrick_2017}, which is the loss of previously learned information when new information is incorporated into the network. In the context of this model, this corresponds to the replacement or destruction of old assemblies in favor of new ones. Although this behavior was not formally addressed in the present work, it is present in the Assembly Calculus model, and overcoming it is crucial for its practical application.
\section*{Conclusions}

We present here an adaptation of an artificial neural network model, called Assembly Calculus, capable of forming neural assemblies based on Hebb’s hypothesis. In this model, the original selection process of which neurons fire at each iteration was replaced by a biologically plausible mechanism based on how excitatory neurons are selected during each gamma oscillation cycle. Additionally, an inhibition method grounded in the excitation-inhibition balance observed in cortical networks was proposed and implemented, making the assembly formation process more controlled. Although the E\%-WTA model exhibited some limitations, it was able to store and represent stimuli within the network for the defined parameters, modulating the size of neural assemblies and enabling robust retrieval of these previously formed neural groups. Finally, to enhance the model’s applicability, other aspects need to be explored with this new approach, such as the storage of stimulus sequences and the retrieval of multiple assemblies stored within the same memory area — an issue related to catastrophic forgetting, which is common in neural networks.

\bibliography{reference}
\bibliographystyle{IEEEtran}
\end{multicols}

\end{document}